\documentstyle[12pt,epsf]{article}
\newcommand{\be}{\begin{equation}}
\newcommand{\ee}{\end{equation}}
\newcommand{\bea}{\begin{eqnarray}}
\newcommand{\eea}{\end{eqnarray}}
\newcommand{\ci}{\cite}
\newcommand{\bi}{\bibitem}

\newcommand{\dd}{\partial}

\newcommand{\half}{\frac{1}{2}}

\def\dal{\,\lower0.3ex\vbox{\hrule\hbox{\vrule\kern2pt\vbox{\kern4pt\kern4pt}
\kern2pt\vrule}\hrule}\,}

\def\o{\omega}

\begin{document}
\title{{\sl Decay to bound states of a soliton in a well}}
\vspace{1 true cm}
\author{G. K\"albermann\\
Faculty of Agriculture,\\ 
Hebrew University, Rehovot 76100,
Israel\\}

\maketitle

\begin{abstract}
\baselineskip 1.5 pc
The decay of a soliton in a trapped state inside a well is shown
numerically. Bound states of a kink in an attractive well,
both centered and off center are found. Their stability is studied.
Unstable soliton solutions inside a repulsive barrier are also
found.
\end{abstract}

%\newpage
%\tableofcontents
\newpage
\baselineskip 1.5 pc

\section{\label{decay}\sl Decay of a soliton in a well}

Solitons, originally observed by Scott Russel in a channel near Edinburgh 
around a century and a half ago and termed by him 'translational
waves', remained as a curiosity for a long time. The past
half century has witnessed the revival of the topic both theoretically
and experimentally. Solitons are now generated in nonlinear transmission lines,
optical fibers (bright and dark solitons), Josephson junctions, 
crystals, etc.\ci{rem}

Almost a decade ago Kivshar et al\ci{kivshar} noted 
that a soliton interacting with an impurity modelled as an external
potential or a space dependent mass term, generates a wealth of
phenomena ranging from repulsion by an attractive impurity and trapping
of a positive kinetic energy soliton in an attractive well.
The mechanism of trapping consists in a soliton arriving at the site of an
attractive potential and oscillating inside it for an infinite time
without the possiblity of reemerging from it. These features were
demonstrated for a spike-like impurity, $\delta$ function, and later
generalized to the case of a finite size potential.\ci{kal97}
(see also ref.~\ci{cuba,cuba1})

Trapping can be understood in terms of a few degrees of freedom
for the soliton\ci{kivshar}, namely, the center of the soliton, 
a shape mode excitation and an impurity mode.\ci{kal99}
These characteristics of soliton scattering from an attractive well
are generic and apply also to Sine-Gordon solitons and perhaps
even other solitons, not researched until now.

The introduction of a potential 
is intended to mock-up the behavior of impurities existing in the path
of the soliton as in large Josephson junctions where
Sine-Gordon solitons are generated. Potentials are real obstacles
from which a soliton scatters.

If a soliton is trapped in a well, the oscillations of the soliton are
a source of radiation that leads to the decay to lower and lower energy states,
until the soliton ends up in a bound state.

The existence of these bound states is here demonstrated for kinks.
Similar states occur in sine-Gordon solitons.
These findings are relevant not only for fluxons, but
for any kind of soliton propagating in an inhomogeneus medium that
can be represented in terms of a potential. 
The equivalence of an inhomogeneous medium and an
impurity potential is demonstrated in ref.\ci{kal99}.

We here proceed to show that a soliton in a well does indeed
decay, its energy being radiated away. We will then find the
final states after the kinetic energy is radiated, which are
true bound states, similar to those found in quantum mechanics.
We will later investigate the stability of the states 
including also the unstable static solutions of a soliton 
inside a repulsive barrier.

The kink lagrangian with a potential impurity is\ci{kivshar,kal97}

\be\label{lag1}
 {\cal L}  = \half \dd_{\mu}\phi\,\dd^{\mu}\phi-
{1 \over 4 }\Lambda~{\bigg(\phi^2 - {m^2\over \lambda}\bigg)}^{2}
\ee
Here 
\be \label{Lambda}
\Lambda = \lambda + V(x)
\ee
$\lambda$ being a constant, and $V(x)$ the impurity potential\ci{kal97}
\be \label{u}
V(x) = h~{cosh^{-2}\bigg({a~(x-x_c)}\bigg)}
\ee
Where $a=\frac{6}{w}$, $w$ being the approximate width of the potential,
and we have chosen a bell-shaped potential for the
sake of exemplification.

Independently of the choice of parameters it is found that
trapped states decay. 
There are many initial configurations (initial impinging velocities)
of the soliton far away from the potential that lead
to trapping \ci{kivshar, kal97}
After the soliton is trapped inside the well, it oscillates
back and forth. Its kinetic energy is still positive.

If the soliton behaved as a pointlike object, it would emerge from the well
unscathed. However, due to its extended nature \ci{kalnew}
it may be trapped. 

The nonlinear wave equation that
determines the solitons has also small amplitude solutions, 
the mesons of the theory\ci{raj}.

These excitations are generated by the oscillations of the soliton
in the well, analogously to an oscillating electric dipole.

The waves are emitted in both directions around
the well location and drain the kinetic energy of the soliton.

Figure 1 shows the amplitude of the oscillation of the soliton,
namely, the value of the field at the center of the well,
as a function of time. Here we used
$m=1,~\lambda=1,~h=-3,~a=2,~x_1=3~$. 
The soliton impinges from the left. The initial location of the 
center of the soliton is chosen to be far enough 
from the well at $x=-3$, with an initial speed
$v=.025$. The soliton is trapped and immediately starts radiating its
energy.

In order to visualize the decay and emission of
radiation we extended the x-axis to $-140~\leq x~\leq 140$ with 
a grid of dx= 0.1. This coordinate span allows for
radiation to progate for a long distance away from the
trapping zone without being reflected.
\begin{figure}
\epsffile{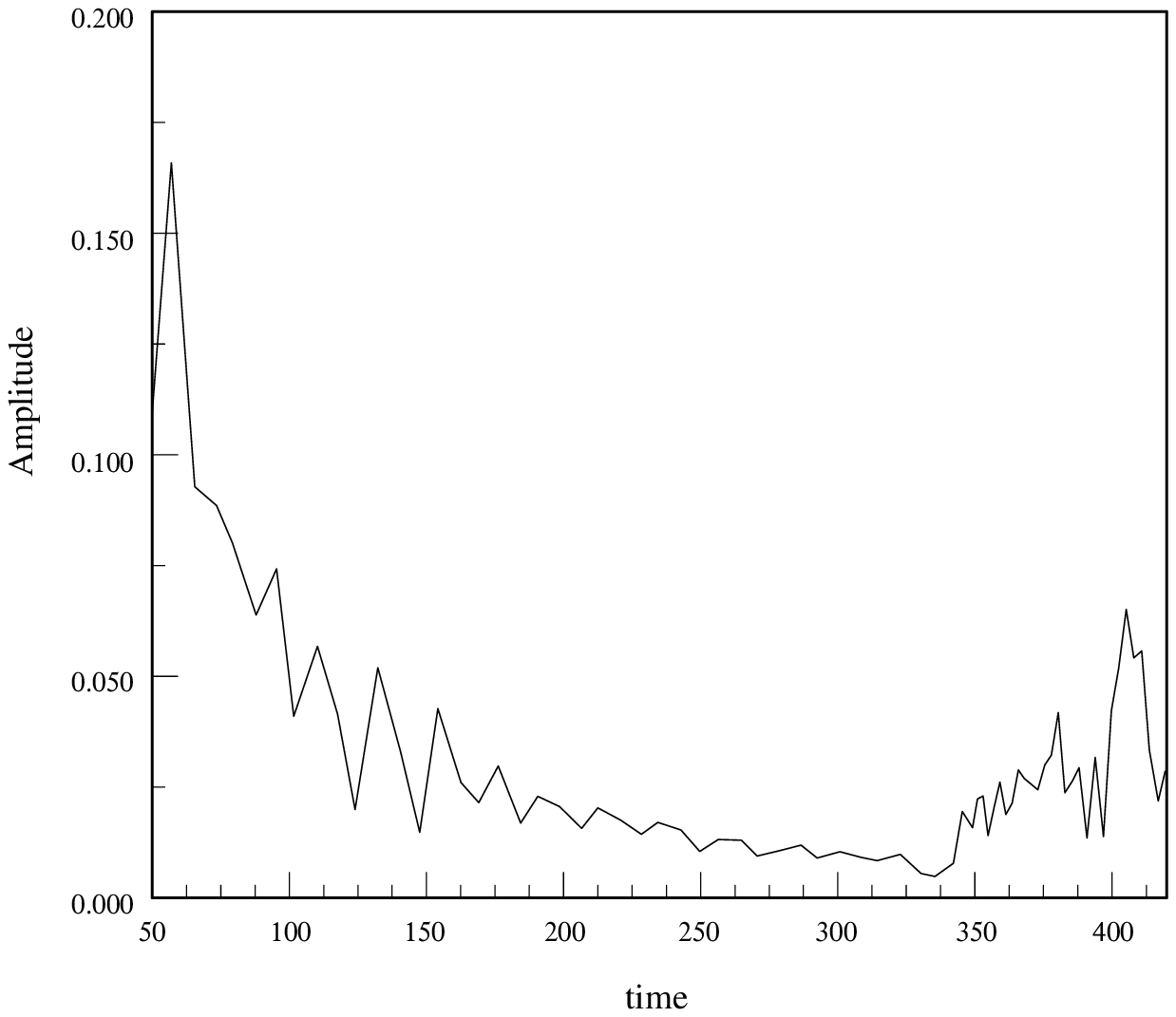}
\caption{Amplitude of the oscillation of the soliton in a trapped
state as a function of time. Soliton parameters: $m=1,~\lambda=1$, 
impurity parameters: $h=-3,~a=2,~x_c=3$.
}
\label{fig1}
\end{figure}

After a certain time, and due to
the finite extent of the x-axis, radiation reflects back
from the boundaries and reaches the soliton. The soliton
subsequently absorbs the radiation and its amplitude starts to
increase.
The time taken for radiation to return to the soliton
is the travel time for the fastest `mesons' of the theory.

The dispersion relation for the radiated mesons can be extracted
from the expansion of the scalar field around the soliton solution. 
Using $\lambda~=~m~=1$ we find $\o^2=k^2+2$. 
The velocity of
the mesons is bounded by 

$u_{max}=\frac{\o}{k}\bigg)_{max} = 1$. This
is clearly observed in figure 1. The reabsorption of radiation starts
after the first mesons arrive back from the boundaries
to the well.
The distance between the well and the boundary
is 140, therefore $t_{absorption}=280/u_{max}=280$

The frequency of the oscillations
of the soliton in a trapped state may be estimated analytically.
Using an expansion of the potential in eq.~(\ref{u})
around the bottom of the well
$V(x)\approx-V_0+\epsilon~y^2,~ y=x-x_c$ and an ansatz appropriate for
small oscillations of the soliton around the center of the well
$\phi\approx(y+\delta~y^3/2)~sin(\o(t-t_0))$ we find $\o^2=2~\mu$.

With $\mu\approx\frac{+}{}\sqrt{~\frac{4}{5}\epsilon} + \frac{9}{10}(V_0-1)$.
(The positive solution has to be chosen)

The formula compares reasonably well with the leading frequency
of oscillation of the soliton inferred from a Fourier analysis of
the amplitude of the field at the center of the well.
However, the fluctuation of the soliton in the well
is anharmonic.

\begin{figure}
\epsffile{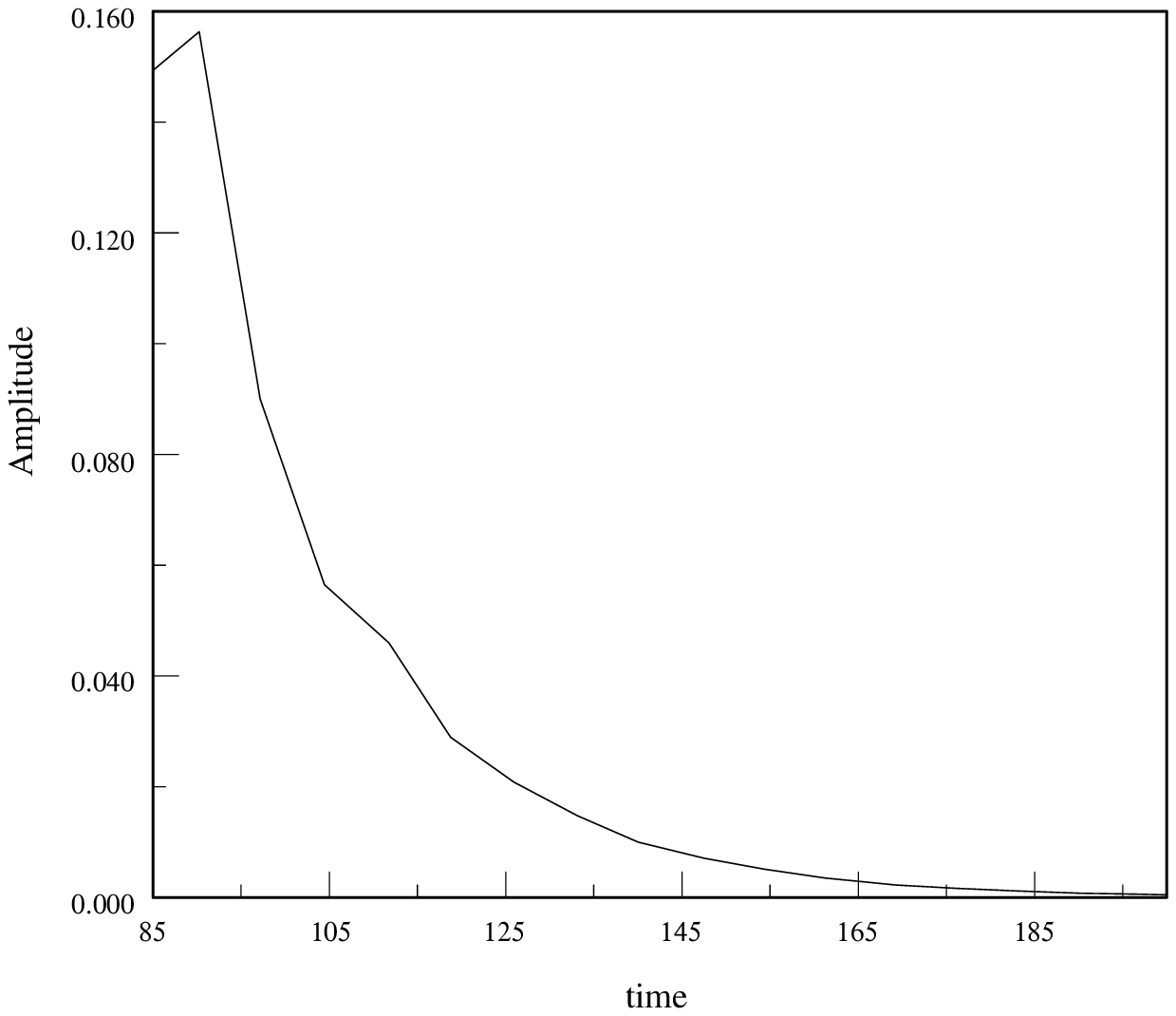}
\caption{Same as figure 1 but including attenuation. Friction
coefficient $\gamma~=~.1$.}
\label{fig2}
\end{figure}

Another way to observe the decay of a trapped state to a bound state
consists of
adding a dissipative force of the form $\gamma \frac{\dd\phi}{\dd t}$.
This force cannot be derived from a Hamiltonian, but, it can arise
from the interaction to an environment.

Inserting this term in the soliton equation of motion yields the 
results depicted in figure 2, where we took 
the same set of  parameters as those of
the radiation run of figure 1, but with a friction coefficient $\gamma=.1$.

Attenuation is the dominant effect in this case. The soliton loses
its energy by dissipation instead of radiating it.
Other choices of parameters may lead to a mixture of both processes.
It is, however, evident that
trapped states will eventually become bound states.
\newpage
\section{\label{states}{\sl Bound states of a soliton in an attractive well}}

In the previous section, it was shown that solitons radiate their energy
when trapped. Hence, there should exist static bound state solutions of
the soliton in the well. 

These bound states exist not only for a soliton centered with the well,
but also for a soliton located off-center.
The former are produced after the soliton radiates its kinetic
energy, while the latter seem more difficult to realize.
This phenomenon has no counterpart in 
the classical behavior of particles. Only the bottom of the well is
the point where the particles can remain motionless. 
The extended character of the soliton is playing a major role
in generating such unexpected solutions.

It is clear that these off-center solutions are
true bound states, because their energy is smaller than the free
soliton mass. However, any small perturbation of the
soliton will make it drift to the center of the well. The
off-center solutions are unstable.

We found the bound state solutions, by integrating the
static equations of motion starting from the center of the 
soliton. There appears to be only a single bound state for
each choice of well depth and width, even for large well depths.

Figure 3 shows a bound state centered with the well,
for a well depths $h=-1$, (solid line), $h=-5$ (dashed line) and a width
parameter $w=5$. Figure 4 shows a solution for an off-center soliton
for a well that is located at a position $x=5$ and the same parameters.
We took wells much wider than the extent of the soliton for
the graphs. For narrower wells, we bring the calculated masses only
in a table below.

\begin{figure}
\epsffile{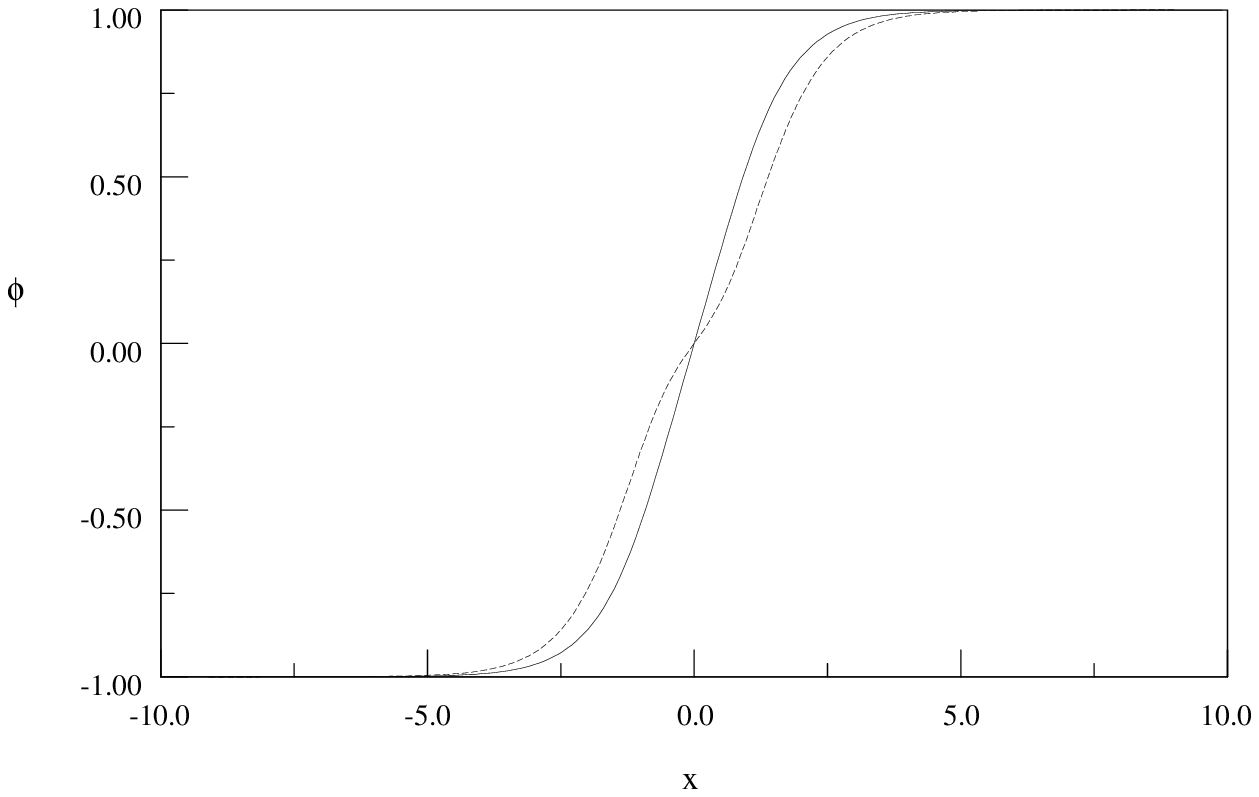}
\caption{Bound state soliton solution 
in a well with parameters h=-1
and w=5, full line and h=-5 w=5, dashed line.}
\label{fig3}
\end{figure}

The deeper the well, the more dramatic the distortion of the soliton
from its free shape.
When the depth of the well exceeds, in absolute value,
the mass parameter entering the soliton equation of motion, 
a topological solution with the right asymptotics 
is obtained by having a vanishing soliton at the location of the well
both for centered and off-center wells as evidenced in figure 4. 
For this reason,
the numerical solutions are extremely sensitive
to the slope of the soliton at the well.
 
Table 1 shows the masses of the soliton for several choices
of depths and widths. These have to be compared
with the free soliton mass of M=0.9428.
The soliton energy may even become negative for deep wells.
A narrow and thin well barely changes the soliton energy. 
The deeper and wider the well, the stronger the effect, as seen from the
last five entries in the table. This feature resembles bound state
solutions of the Schr\"odinger equation.

\begin{figure}
\epsffile{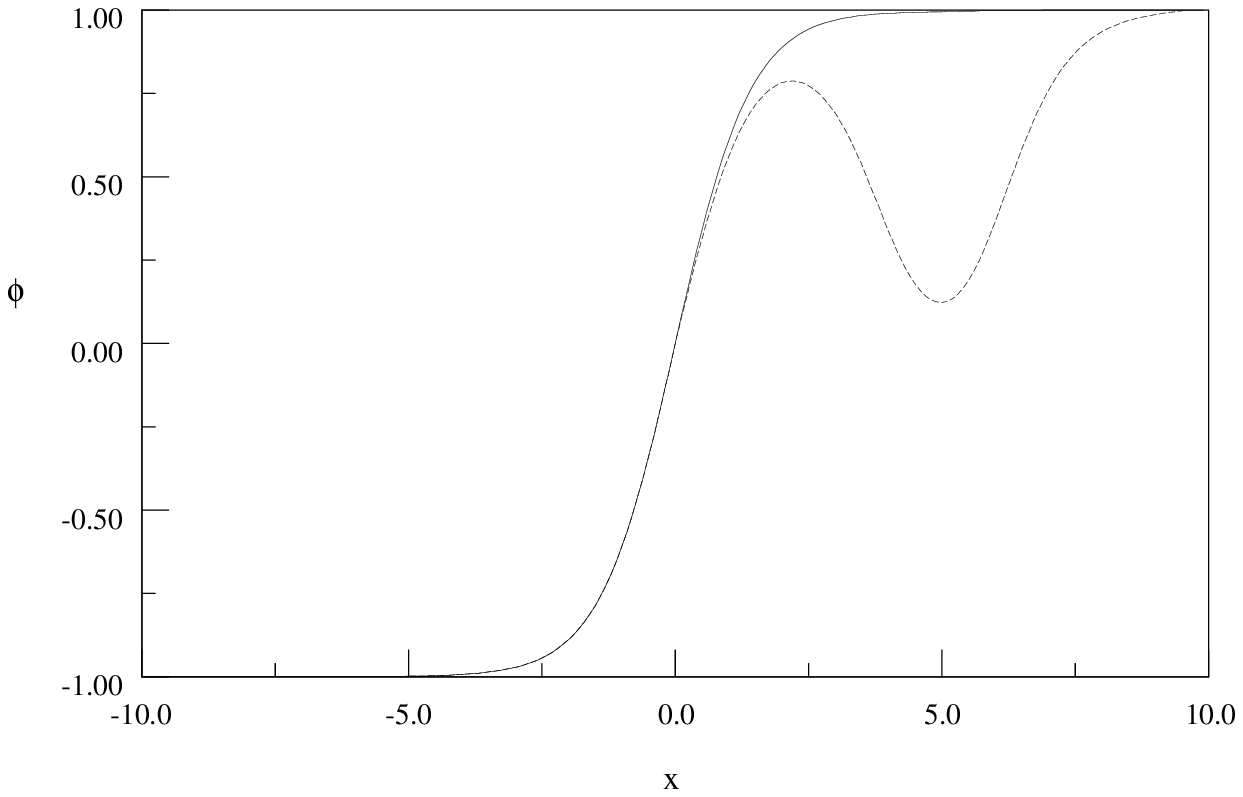}
\caption{Same as figure 3, but for a well located at x=5}
\label{fig4}
\end{figure}

\begin{table}
\caption {{\bf Total energy of a soliton in a well}}
	\begin{center}
         \medskip
         \begin{tabular}{|c|c|c|c|}
           \hline
well depth&width parameter {\sl w}&well center&soliton energy\\
&&&\\
           \hline
-1.&5&0&0.633\\
&&1&0.7178\\
&&3&0.9308\\
&&5&0.9425\\
           \hline
-5&1&0&0.533\\
&&1&0.676\\
&&3&0.9312\\
&&5&0.9425\\
           \hline
-5&5&0&-0.771\\
&&1&-0.6834\\
&&2&-0.4638\\
&&4&0.025\\
&&5&0.1177\\
           \hline
         \end{tabular}
     \end{center}
\end{table}
\newpage

Let us now consider the stability of the soliton solutions 
for the centered case. We do not have analytical expressions
for the soliton, therefore we proceed by considering the
behavior of the soliton around its center.

In a small region around the bottom of the well,
we can write the soliton solution in terms of a new mass
parameter $\Lambda=1+ V(x)$, for $x\approx 0$,
as $\phi~=\tanh\bigg(\sqrt{\frac{\Lambda}{2}}~(x-x_0)\bigg)$, 
where $x_0$ is the location of the
soliton that we deliberately move away by a small amount from $x=0$.
In the above ansatz we have used the mass and width parameters as
previously $m=1,~\lambda=1$.
For $-V(0)>1$ we use the $\sl tan$ solution instead of the
$\sl tanh $  one.
With such an ansatz in the soliton equation of motion we find
the equation for the center of the soliton

\be\label{osc}
\ddot{x}_0=~\frac{\frac{+}{}~2}{\Lambda(x_0)}~V'(x_0)
\ee

where, the upper sign corresponds to a deep well for which the
depth is greater than the mass parameter, $\Lambda(x=0)<0$, and viceversa.
The dots represent time derivatives, the prime space derivative, and
velocities are unitless.
From equation \ref{osc}, it is seen that an attractive well, 
the soliton will be dragged back to the center of the well.
The soliton solutions at the center of the well are stable.
Using similar techniques , we can show that the off-center solitons are
unstable. In this case,
a suitable ansatz for the soliton around the center of the well
for distances not too far from the center of the soliton,
is a quadratic function of the form 
$\phi=A\bigg(1~-\frac{\Lambda}{2}(x-x_0)^2\bigg)$
with $\sl A$, a constant.
We obtain the same equation \ref{osc} but with a reverse sign.
The soliton becomes unstable.
It is clear that small amplitude excursions of the soliton
from its bound state do not kick it out of the well.
However, large amplitude oscillations (maybe
produced by radiation) are able to do so. This is essentially
the time reversed version of the trapping mechanism.

Finally let us note that there exist static soliton solutions
for a repulsive well, with a soliton located at the center of the
well. Due to the convexity of the barrier around its maximum,
the soliton is unstable here also.
Figure 5 shows one such solution for parameters h=1 w=5.
\begin{figure}
\epsffile{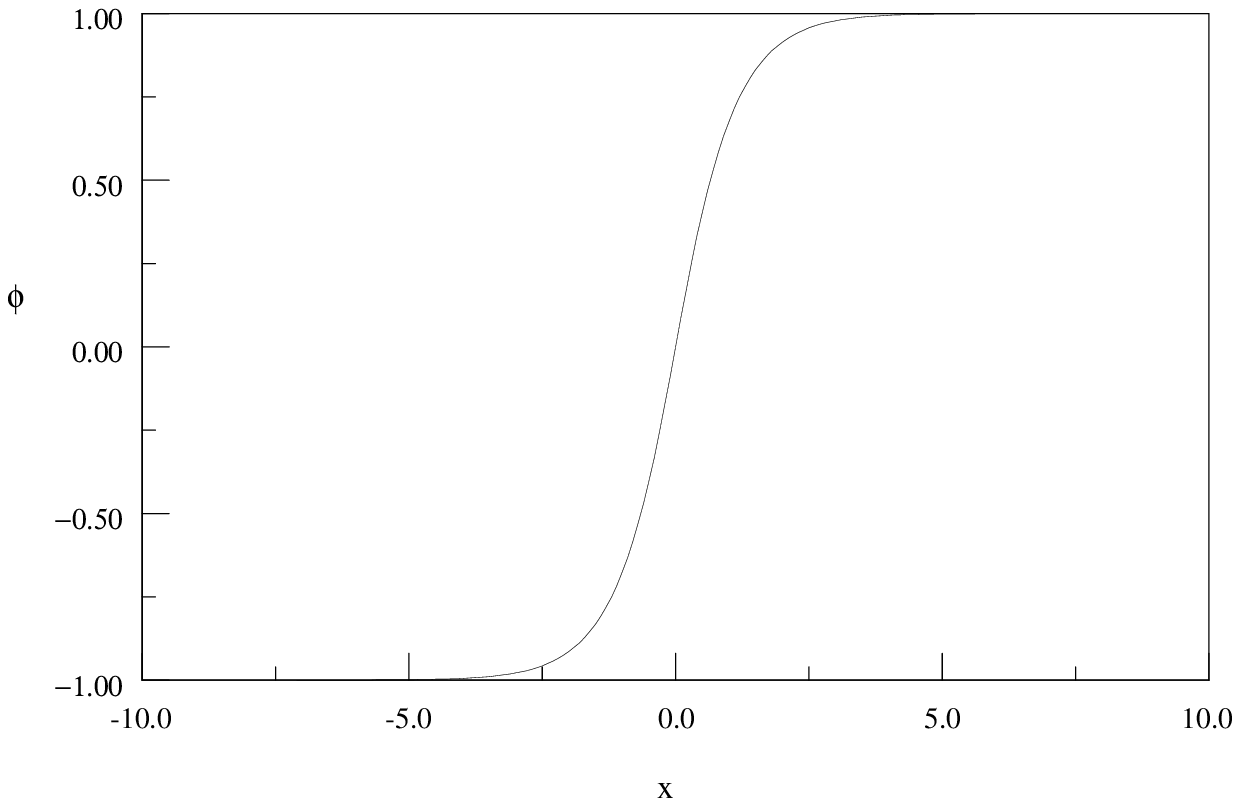}
\caption{Soliton solution 
inside a barrier with parameters h=1
and a=5}
\label{fig5}
\end{figure}

In summary, if a soliton is trapped in an attractive well it will
decay to a stable bound state. It remains for experiments
to find the right conditions for the process to happen. 

{\bf Acknowledgements}

This work was supported in part by the Department of
Energy under grant DE-FG03-93ER40773 and by the National Science Foundation
under grant PHY-9413872.

\end{document}